\documentclass[letter, 10pt, conference]{IEEEtran}

\usepackage[T1]{fontenc}

\usepackage[T1]{fontenc}

\newcommand{\ie}{i.\,e.\ }

\usepackage{tikz, tikz-3dplot}
\usetikzlibrary{positioning, patterns}
\usetikzlibrary{arrows,automata, shapes, petri}
\usetikzlibrary{arrows.meta, shapes.callouts, math}
\usetikzlibrary{calc,decorations.pathreplacing}
\usepackage{tikz-timing}
\usepackage{pgfplots}
\usetikzlibrary{fadings}

\tikzset{router/.style={circle,draw,fill=gray!60,inner sep=0pt,minimum size=5pt}}
\tikzset{desc/.style={font = \scriptsize}}
\tikzset{cross/.style={cross out, draw, 
		minimum size=2*(#1-\pgflinewidth), 
		inner sep=0pt, outer sep=0pt}}

\definecolor{col1}{RGB}{27,158,119}
\definecolor{col2}{RGB}{217,95,2}
\definecolor{col3}{RGB}{117,112,179}

\usepackage{subcaption}
\usepackage{amsmath}

\usepackage{units}
\usepackage[hyphens]{url}
\usepackage{booktabs}

\usepackage{tabularx}
\newcolumntype{Z}{>{\raggedright\let\newline\\\arraybackslash}X}
\usepackage{multirow}
\usepackage{rotating}

\tikzstyle{router}=[draw,
fill=black!25,
circle,
minimum size=10pt,
inner sep=0pt,
font=\scriptsize]
	
\newcommand{\myGlobalTransformation}[3]
{
    \pgftransformcm{1}{0}{0.4}{0.35}{\pgfpoint{#1cm}{#2cm}}
    \pgftransformscale{#3}
}

\newcommand{\drawNoC}[4]
{
    \pgfmathtruncatemacro{\xNodes}{#1};
    \pgfmathtruncatemacro{\yNodes}{#2};
    \pgfmathtruncatemacro{\zNodes}{#3};
    
    \ifthenelse{\equal{\zNodes}{1}}{
        \pgftransformscale{\yNodes*0.8}
        \fill [black!5, draw=black, line width=0.2mm] (-0.2cm,-0.2cm )rectangle (1.2cm,1.2cm);

        \foreach \x in {1,...,\xNodes}{
            \foreach \y in {1,...,\yNodes} {
                \pgfmathtruncatemacro{\n}{(\y-1)*#1+(\x-1)};
                    \node[circle, draw=black, fill=white, line width=0.2mm, inner sep=0pt,minimum size=14pt, font=\bfseries, nodeStyle\n/.try] (\n) at ({(\x-1)/(\xNodes-1)},{(\y-1)/(\yNodes-1)}) {\tiny \n};
            }
        }
        \def\temp{#4}\ifx\temp\empty
        \else
            \node at (0.5,-0.3) {#4};
        \fi
    
    }
    {
        \foreach \z in {1,...,\zNodes}{
            \myGlobalTransformation{0}{(\z-1)*\yNodes*0.625}{\yNodes};
                \fill [black!5, draw=black, line width=0.2mm] (-0.2cm,-0.3cm )rectangle (1.3cm,1.3cm);

                \foreach \x in {1,...,\xNodes}{
                    \foreach \y in {1,...,\yNodes} {
                        \pgfmathtruncatemacro{\n}{(\y-1)*#1+(\x-1)+((\z-1)*\xNodes*\yNodes)};
                    \node[circle, draw=black, fill=white, line width=0.2mm, inner sep=0pt,minimum size=11pt, font=\bfseries, nodeStyle\n/.try] (\n) at ({(\x-1)/(\xNodes-1)},{(\y-1)/(\yNodes-1)}) {\tiny \n};
            }
        }

    }  
        \def\temp{#4}\ifx\temp\empty
        \else
            \node at (1.5,-0.65) {#4};
        \fi
    }

    \setOffset{0};
}

\newcommand{\drawNoCNumbered}[5]
{
	\pgfmathtruncatemacro{\xNodes}{#1};
	\pgfmathtruncatemacro{\yNodes}{#2};
	\pgfmathtruncatemacro{\zNodes}{#3};
	
	\ifthenelse{\equal{\zNodes}{1}}{
		\pgftransformscale{\yNodes*0.8}
		\fill [black!5, draw=black, line width=0.2mm] (-0.2cm,-0.2cm )rectangle (1.2cm,1.2cm);
		
		\foreach \x in {1,...,\xNodes}{
			\foreach \y in {1,...,\yNodes} {
				\pgfmathtruncatemacro{\n}{(\y-1)*#1+(\x-1)};
				\ifthenelse{\equal{#5}{1}}{
					\node[router, nodeStyle\n/.try] (\n) at ({(\x-1)/(\xNodes-1)},{(\y-1)/(\yNodes-1)}) {\tiny \n};
				}
				{
					\node[router, nodeStyle\n/.try] (\n) at ({(\x-1)/(\xNodes-1)},{(\y-1)/(\yNodes-1)}) {};
				}
			}
		}
		\def\temp{#4}\ifx\temp\empty
		\else
		\node at (0.5,-0.3) {#4};
		\fi
		
	}
	{
		\foreach \z in {1,...,\zNodes}{
			\myGlobalTransformation{0}{(\z-1)*\yNodes*0.625}{\yNodes};
			\fill [black!5, draw=black, line width=0.2mm] (-0.2cm,-0.3cm )rectangle (1.3cm,1.3cm);
			
			\foreach \x in {1,...,\xNodes}{
				\foreach \y in {1,...,\yNodes} {
					\pgfmathtruncatemacro{\n}{(\y-1)*#1+(\x-1)+((\z-1)*\xNodes*\yNodes)};
					\ifthenelse{\equal{#5}{1}}{
						\node[router, nodeStyle\n/.try] (\n) at ({(\x-1)/(\xNodes-1)},{(\y-1)/(\yNodes-1)}) {\tiny \n};
					}
					{
						\node[router, nodeStyle\n/.try] (\n) at ({(\x-1)/(\xNodes-1)},{(\y-1)/(\yNodes-1)}) {};
					}
				}
			}
			
		}  
		\def\temp{#4}\ifx\temp\empty
		\else
		\node at (1.5,-0.65) {#4};
		\fi
	}

	\setOffset{0};
}

\newcommand{\drawANoC}[4]
{
	\pgfmathtruncatemacro{\xNodes}{#1};
	\pgfmathtruncatemacro{\yNodes}{#2};
	\pgfmathtruncatemacro{\zNodes}{#3};
	
	\ifthenelse{\equal{\zNodes}{1}}{
		\pgftransformscale{\yNodes*0.8}
		\fill [black!5, draw=black, line width=0.2mm] (-0.2cm,-0.2cm )rectangle (1.2cm,1.2cm);
		
		\foreach \x in {1,...,\xNodes}{
			\foreach \y in {1,...,\yNodes} {
				\pgfmathtruncatemacro{\n}{(\y-1)*#1+(\x-1)};
				\node[circle, draw=black, fill=white, line width=0.2mm, inner sep=0pt,minimum size=14pt, font=\bfseries, nodeStyle\n/.try] (\n) at ({(\x-1)/(\xNodes-1)},{(\y-1)/(\yNodes-1)}) {\tiny \n};
			}
		}
		\def\temp{#4}\ifx\temp\empty
		\else
		\node at (0.5,-0.3) {#4};
		\fi
		
	}
	{
		\foreach \z in {1,...,\zNodes}{
			\myGlobalTransformation{0}{(\z-1)*\yNodes*0.625}{\yNodes};
			\ifthenelse{\equal{\z}{1}}{
				\fill [black!5, draw=black, line width=0.2mm] (-0.2cm,-0.3cm )rectangle (1.3cm,1.3cm);
			}{
				\fill [black!30, draw=black, line width=0.2mm] (-0.2cm,-0.3cm )rectangle (1.3cm,1.3cm);
			}
			
			\foreach \x in {1,...,\xNodes}{
				\foreach \y in {1,...,\yNodes} {
					\pgfmathtruncatemacro{\n}{(\y-1)*#1+(\x-1)+((\z-1)*\xNodes*\yNodes)};
						\ifthenelse{\equal{\z}{1}}{
							\node[circle, draw=black, fill=white, line width=0.2mm, inner sep=0pt,minimum size=11pt, font=\bfseries, nodeStyle\n/.try] (\n) at ({(\x-1)/(\xNodes-1)},{(\y-1)/(\yNodes-1)}) {\tiny \n};
						}{
							\node[circle, draw=black, fill=lightgray, line width=0.2mm, inner sep=0pt,minimum size=11pt, font=\bfseries, nodeStyle\n/.try] (\n) at ({(\x-1)/(\xNodes-1)},{(\y-1)/(\yNodes-1)}) {\tiny \n};
						}
				}
			}
			
		}  
		\def\temp{#4}\ifx\temp\empty
		\else
		\node at (1.5,-0.65) {#4};
		\fi
	}

	\setOffset{0};
}

\newcommand{\drawNoCTwo}[5]
{
    \pgfmathtruncatemacro{\xNodes}{#1};
    \pgfmathtruncatemacro{\yNodes}{#2};
    \pgfmathtruncatemacro{\zNodes}{#3};
    
    \ifthenelse{\equal{\zNodes}{1}}{
        \pgftransformscale{\yNodes*0.8}
        \fill [black!5, draw=black, line width=0.2mm] (-0.2cm,-0.2cm )rectangle (1.2cm,1.2cm);

        \foreach \x in {1,...,\xNodes}{
            \foreach \y in {1,...,\yNodes} {
                \pgfmathtruncatemacro{\n}{(\y-1)*#1+(\x-1)};
                    \node[circle, draw=black, fill=white, line width=0.2mm, inner sep=0pt,minimum size=14pt, font=\bfseries, nodeStyle\n/.try] (\n) at ({(\x-1)/(\xNodes-1)},{(\y-1)/(\yNodes-1)}) {\tiny \n};
            }
        }
        \def\temp{#4}\ifx\temp\empty
        \else
            \node at (0.5,-0.3) {#4};
        \fi
    
    }
    {
        \foreach \z in {1,...,\zNodes}{
            \myGlobalTransformation{0}{((\z-1)*#5)*\yNodes*0.625}{\yNodes};

                \fill [black!5, draw=black, line width=0.2mm] (-0.2cm,-0.3cm )rectangle (1.2cm,1.3cm);

                \foreach \x in {1,...,\xNodes}{
                    \foreach \y in {1,...,\yNodes} {
                        \pgfmathtruncatemacro{\n}{(\y-1)*#1+(\x-1)+((\z-1)*\xNodes*\yNodes)};
                    \node[circle, draw=black, fill=white, line width=0.2mm, inner sep=0pt,minimum size=11pt, font=\bfseries, nodeStyle\n/.try] (\n) at ({(\x-1)/(\xNodes-1)},{(\y-1)/(\yNodes-1)}) {\tiny \n};
            }
            
        }
    }  
        \def\temp{#4}\ifx\temp\empty
        \else
            \node at (1.5,-0.65) {#4};
        \fi
    }

    \setOffset{0};
}

\newcommand{\putNodesBetween}[7]
{
    \pgfmathtruncatemacro{\xNodes}{#1-1};
    \pgfmathtruncatemacro{\yNodes}{#2-1};
    \pgfmathtruncatemacro{\zNodes}{#3-1};
    
    \ifthenelse{\equal{\zNodes}{0}}{
        \foreach \x in {1,...,\xNodes}{
            \foreach \y in {1,...,\yNodes} {
                \pgfmathtruncatemacro{\n}{(\y-1)*\xNodes+(\x-1)};
                \pgfmathtruncatemacro{\nbl}{(\y-1)*\xNodes+(\x-1)+\y-1};
                \pgfmathtruncatemacro{\ntr}{(\y)*\xNodes+(\x-1)+\y+1};

                \node[circle, draw=black, fill=white, line width=0.2mm, inner sep=0pt,minimum size=14pt, font=\bfseries, #5, nodeStyle#4\n/.try] (#4\n) at ($(\nbl)!0.5!(\ntr)$) {\tiny \n};
                
                \ifthenelse{\equal{#6}{1}}{
                    \pgfmathtruncatemacro{\nbr}{(\y-1)*\xNodes+(\x-1)+\y};
                    \pgfmathtruncatemacro{\ntl}{(\y)*\xNodes+(\x-1)+\y};
                    
                    \draw[#7] (\nbl)--(#4\n);
                    \draw[#7] (\nbr)--(#4\n);
                    \draw[#7] (\ntl)--(#4\n);
                    \draw[#7] (\ntr)--(#4\n);
                    \ifnum\y>1%
                        \pgfmathtruncatemacro{\nb}{(\y-2)*\xNodes+(\x-1)};
                        \draw[#7] (#4\n)--(#4\nb);
                    \fi
                    
                    \ifnum\x>1%
                        \pgfmathtruncatemacro{\nl}{(\y-1)*\xNodes+(\x-2)};
                        \draw[#7] (#4\n)--(#4\nl);
                    \fi
                }
            }
        }
    }
    {

    }

    \setOffset{0};
}

\newcommand{\setOffset}[1]
{
    \pgfmathtruncatemacro{\offset}{#1};
}
\definecolor{MyLightGreen}{HTML}{66CC33}
\definecolor{MyPurple}{HTML}{9900CC}
\definecolor{MyLightBlue}{HTML}{0066FF}
\definecolor{tablegray}{rgb}{0.95,.95,.95}
\newcommand*\circled[1]{\tikz[baseline=(char.base)]{
            \node[shape=circle,draw,inner sep=2pt] (char) {#1};}}

\usepackage{csvsimple}
\usepackage{multicol}
\usepackage{enumitem}
\usepackage{colortbl}
\usepackage{lscape}
\usepackage{adjustbox}

\usepackage{tikz-timing}
\usetikztiminglibrary[new={char=Q,reset char=R}]{counters}

\setlist[description]{leftmargin=\parindent}

\usepackage{booktabs} 

\begin{document}
\title{System-level optimization of Network-on-Chips for heterogeneous 3D System-on-Chips\vspace{-12pt}}

\author{
	\IEEEauthorblockN{{Jan Moritz Joseph}\IEEEauthorrefmark{1}, {Dominik Ermel}\IEEEauthorrefmark{1}, {Lennart Bamberg}\IEEEauthorrefmark{2},  {Alberto Garc\'ia-Oritz}\IEEEauthorrefmark{2}, {Thilo Pionteck}\IEEEauthorrefmark{1}}\\\vspace{-12pt}
	\IEEEauthorblockA{\IEEEauthorrefmark{1}Otto-von-Guericke-Universit\"at Magdeburg,
		Institut f\"ur Informations- und Kommunikationstechnik, Germany\\\vspace{-12pt}
		Email: \{jan.joseph, dominik.ermel, thilo.pionteck\}@ovgu.de}\\
	\IEEEauthorblockA{\IEEEauthorrefmark{2}University of Bremen
		Institute of Electrodynamics and Microelectronics, Germany\\
		Email: \{agarcia, bamberg\}@item.uni-bremen.de}}

\maketitle
\begin{abstract}
	For a system-level design of Networks-on-Chip for 3D heterogeneous System-on-Chip (SoC), the locations of components, routers and vertical links are determined from an application model and technology parameters. In conventional methods, the two inputs are accounted for separately; here, we define an integrated problem that considers both application model and technology parameters. We show that this problem does not allow for exact solution in reasonable time, as common for many design problems. Therefore, we contribute a heuristic by proposing design steps, which are based on separation of intralayer and interlayer communication. The advantage is that this new problem can be solved with well-known methods. We use 3D Vision SoC case studies to quantify the advantages and the practical usability of the proposed optimization approach. We achieve up to 18.8\% reduced white space and up to 12.4\% better network performance in comparison to conventional approaches.
\end{abstract}


\maketitle

\section{Introduction}
\vspace{-2pt}

In heterogeneous 3D integrated System-on-Chips (SoCs), dies in disparate technologies (e.g. mixed-signal and digital nodes) are stacked and vertically connected. This is promising, as components with different requirements to technology can be integrated efficiently. Heterogeneity gains attention in the industry, e.g. Intel's recent architecture "Lakefield" \cite{Intel.2019a}. 

Communication in these SoCs can be realized via Networks-on-chip (NoCs). Heterogeneity has a vast influence on the NoC design, as Ref. \cite{Bamberg.2018} shows at the \emph{physical level} and Ref. \cite{Joseph.2017} at the \emph{architectural level}. At the \emph{system level}, the locations of components (\ie PEs), routers and vertical links, as well as the network topology are optimized. 
Many aspects of the system-level optimization have been solved (e.g. floorplaning \cite{Cong.2004} or NoC topology synthesis \cite{Seiculescu.2010}), but their integrated solution has not been considered sufficiently so far. In heterogeneous 3D SoC the necessity for an integrated approach arises from the varying components' and routers' costs between the layers plus the mutual influence between the vertical interconnect planning and the router / component placement. 

This paper targets such an integrated approach. Three fundamental characteristics of heterogeneity are identified for the system-level optimization (Sec.~\ref{sec:tech}). Next, \emph{the system-level optimization problem is defined} (Sec.~\ref{sec:tech}). As for many design problems, it is impossible to calculate an exact solution in reasonable time. Thus, \emph{we propose five design steps that efficiently generate a solution} by separation of the intralayer and the interlayer communication (Sec.~\ref{sec:heur}). Each individual step is elementary to enable efficient solving; existing methods can be used with small modifications considering the fundamental characteristics. The heuristic is applied to heterogeneous 3D SoC case studies to quantify its advantages (Sec.~\ref{sec:results}).

\begin{figure}
	\begin{minipage}[b]{0.32\linewidth}
		\centering
		\includegraphics[width=.95\linewidth]{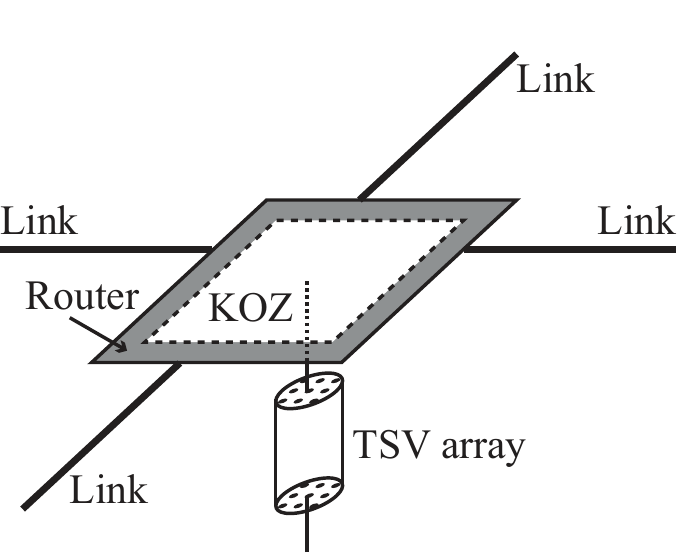}
		\subcaption{Downward.}
		\label{fig:optmization:routermodel:down}
	\end{minipage}\hfill	\begin{minipage}[b]{0.32\linewidth}
		\centering
		\includegraphics[width=.95\linewidth]{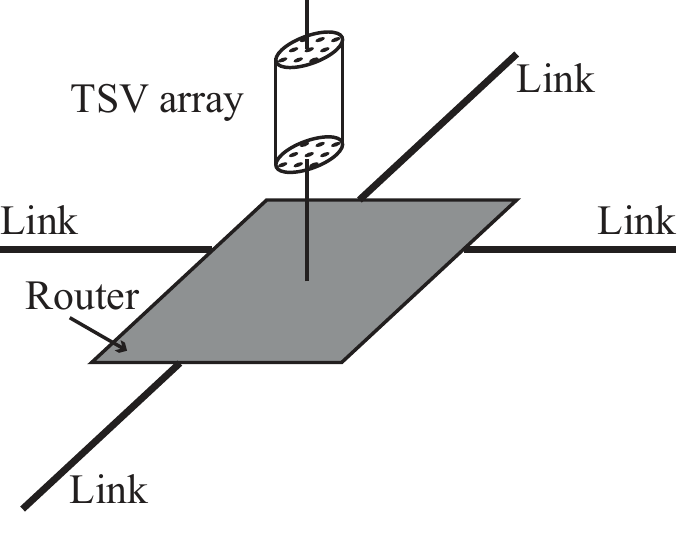}
		\subcaption{Upward.}
		\label{fig:optmization:routermodel:uponly}
	\end{minipage}\hfill
	\begin{minipage}[b]{0.32\linewidth}
		\centering
		\includegraphics[width=.95\linewidth]{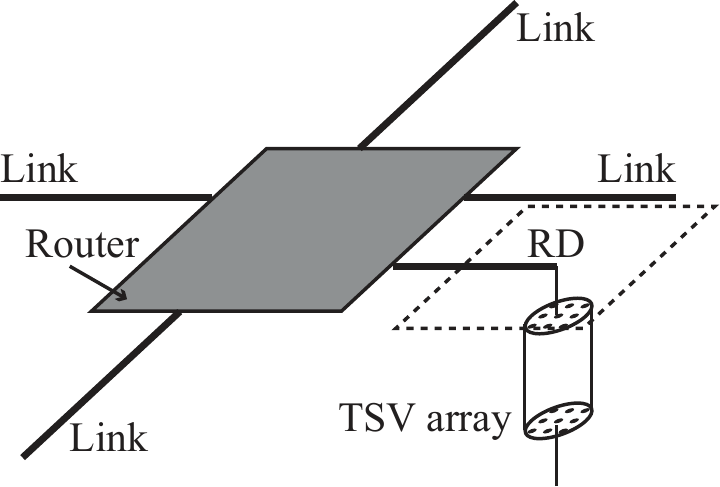}
		\subcaption[Vertical connection with RD]{Redistribution.}
		\label{fig:optmization:routermodel:up}
	\end{minipage}
	\vspace{-6pt}
	\caption[Router model]{Router model with KOZ and RD.}
	\label{fig:optmization:routermodel}\vspace{-10pt}
\end{figure}

\vspace{-2pt}
\section{The Integration Issues}\label{sec:tech}
\vspace{-2pt}


The following three fundamental characteristics of heterogeneous 3D integration influence the system-level optimization:


1) Routers and components have varying performance, power and area (PPA) per technology node / per layer. Some components cannot be implemented in certain nodes due to physical limitations, e.g.\ analog sensors in digital technologies. 

2) Through-Silicon Vias (TSV) arrays are used as vertical interconnects. TSVs are commonly manufactured in the via-middle process flow and thus yield keep-out-zones (KOZs) when crossing layers. Therefore, area must be reserved.

3) It is beneficial to use redistribution (RD) that connects the TSV arrays with the routers via horizontal metal wires. This allows the routers to be vertically connected without being exactly above each other. The length of the RD is limited by the target clock frequency and the technology. 



These characteristics influence the NoC: First, a partially vertically connected mesh topology is more area-efficient than a fully-connected NoC. Second, the KOZs and the RD must be modeled.  If a router connects downwards, there will be a KOZ (Fig.\,\ref{fig:optmization:routermodel:down}). If it connects upwards, there is none (Fig.\,\ref{fig:optmization:routermodel:uponly}). If RD is used, the KOZ will be outside of the router (Fig.\,\ref{fig:optmization:routermodel:up}).


\vspace{-3pt}
\section{The Problem formulation}\label{sec:prob}
\vspace{-2pt}

\begin{figure*}
	\centering
	\includegraphics[width=.85\linewidth]{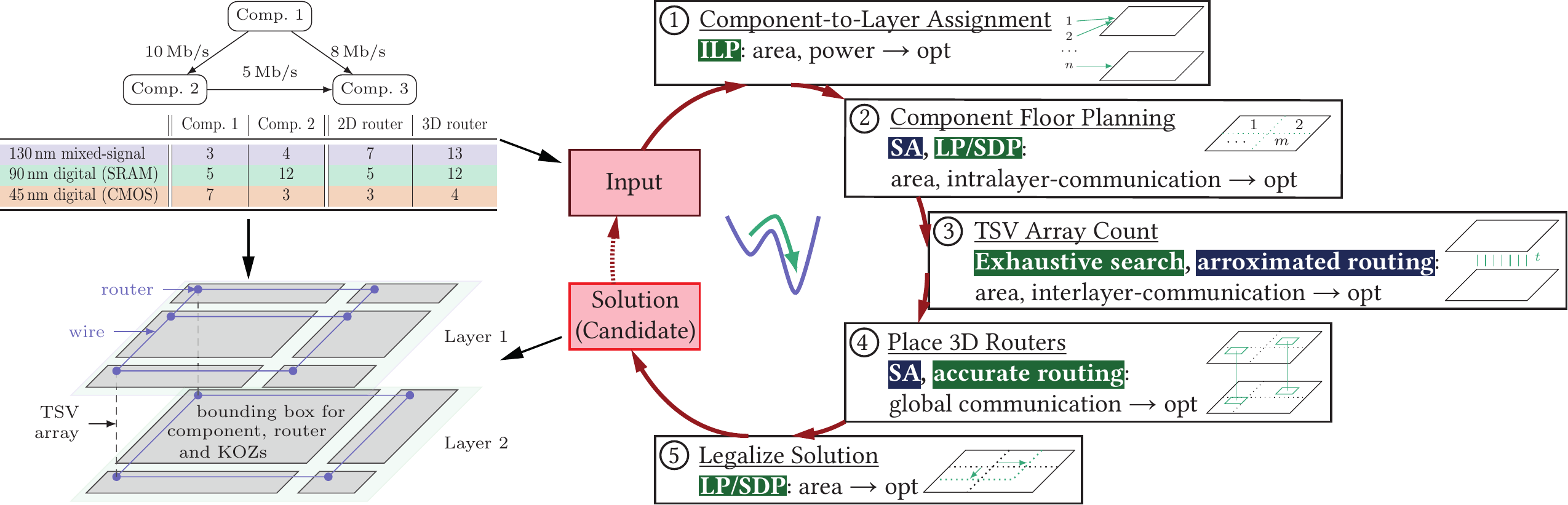}
	\caption{Problem definitions with input and output \& steps of the heuristic (green: optimal solution. blue: heuristic solution).}
	\label{fig:problemcomplete}\vspace{-12pt}
\end{figure*}

We define the system-level optimization of NoCs considering heterogeneous integration. Its input is divided into application and technology properties, shown in Fig.~\ref{fig:problemcomplete}, upper left-hand side: The \emph{application} is described by a core graph \cite{Srinivasan.2006} in that the nodes are the components and the edge weights model the bandwidth between the components. The \emph{technology} properties of the components and routers are given by PPA tables. At the example of area, the table gives $area_{c,l}$, in which $c \in C$ denotes a component $c$ in the set of all components $C$ and $l \in L$ one layer in the set of all layers. Let $C(l)$ be the subset of components in the layer $l$. Plus, the KOZ area $K$ and the RD length $R$ are given. 

The result of the optimization is a network graph, in that the nodes are routers and the edges are links. Plus, rectangular bounding boxes reserve space, each for one component and one 2D router or one 3D router with a KOZ. The bounding boxes follow the mesh topology. Fig.~\ref{fig:problemcomplete} (left-hand side) shows an example with bounding boxes and a network graph. 

\setlength{\abovedisplayskip}{0pt}
\setlength{\belowdisplayskip}{0pt}
\setlength{\abovedisplayshortskip}{0pt}
\setlength{\belowdisplayshortskip}{0pt}

The objective is linear with five weights for scaling: 
\begin{align*}
c = \omega_1 c_{{area}}+ \omega_2 c_{{power}} + \omega_3 c_{{perf}} + \omega_4 c_{{peak}} + \omega_5 c_{{util}}
\end{align*}
The first three addends are the chip area $c_{area}$, the summed (component) performance $c_{perf}$, and the summed power consumption $c_{power}$. This models PPA, but does not account for the network performance. Thus, the fourth addend $c_{peak}$ models the throughput by penalizing congested links with a higher load than available bandwidth. The fifth addend $c_{util}$ models the latency, measured in hop distance $\times$ bandwidth. This linear model allows for extensions, e.g. the standard methods for thermal optimization (e.g. \cite{Cong.2004, Seiculescu.2010}) are also linear. 
\vspace{-8pt}
\section{The Heuristic: Five Design Steps}\label{sec:heur}
\vspace{-2pt}

We define a heuristic for the system-level optimization by the identification of five design steps. Efficient run-time is possible due to separation of interlayer and intralayer communication. The steps are shown in Fig.~\ref{fig:problemcomplete}, right-hand side.

\textbf{Step {\scriptsize\circled{1}}}: The \textbf{component-to-layer assignment} minimizes the weighted sum of the maximum chip area across layers (which dominates the area of the 3D stack) and the total power: $C_1 = w_1 c_{area} + w_2 c_{power} = w_1 \max_{l \in L} \sum_{ c \in C(l)} area_{c, l} + w_2 \sum_{l \in L, c \in C} power_{c, l}$. The problem is formulated as an integer LP (ILP). Since the max-function is minimized, it can be modeled with an auxiliary variable $z$ that is constraint by the max-function's inputs ($z \geq max(a, b) \rightarrow z \geq a, z\geq b$). The ILP is constraint such that each component is assigned to one layer only. 
The layer order is not optimized as this would require to consider the communication as early as in this step. 

\textbf{Step {\scriptsize\circled{2}}}: The \textbf{component floor planning} minimizes the layer area and the interlayer communication. This step can be done individually per layer, as routers must not be placed at the same position due to RD. Per layer, $C_2 = w_1 c_{area} + w_4 c_{peak} + w_5 c_{util} $ is minimized with a simulated annealing (SA). It optimizes the floorplan in a 2D mesh topology by switching the components' assignment to mesh-columns and rows. Here, the component area is not simply added as in step {\scriptsize\circled{1}}; rather, the complete layer's area is minimized for each floorplan in the SA by minimizing the width $\omega_{i, j}$ and length $\lambda_{i,j}$ for all components at columns $i$ and rows $j$ in the 2D mesh. This optimization is constraint by the component areas $area_{c, l}$ at column/row $i,j$: $\omega_{i,j}\lambda_{i,j} \geq area_{c, l}$ for all $i,j$. The KOZ and 3D router area cannot be accounted for, at this time. For the area minimization within the SA, a linear program (LP) and a semi-definite program (SDP) can be used.  In the LP, the area must be approximated as the multiplication $\omega_{i,j}\lambda_{i,j}$ is non-linear, while the SDP is exact, as it is possible to model $\omega_{i,j}\lambda_{i,j}$. For the equations of both LP and SDP as well as a performance analysis, see Ref.~\cite{Joseph.2019}. 

\textbf{Step {\scriptsize\circled{3}}}: The \textbf{TSV array count} minimizes the area and maximizes the interlayer communication, as more TSVs reduce link load but require more KOZs. The exact routing, and thus the communication, is unknown; hence, it is given by the expected bandwidth $b_j$ for the $j$-th TSV array. This is calculated by uniform-randomly distributing TSV arrays to the grid and use the floor plan to find the nearest-neighbored components with the Manhattan distance $d_j$ and sum the component's upward and downward bandwidth. Thus, the objective for $i$ TSVs is $C_3 =iK + \sum_{j = 1}^{i}{b_jd_j}$. As the TSV count is limited, exhaustive search finds the global minimum. 

\textbf{Step {\scriptsize\circled{4}}}: The \textbf{placement of 3D routers} minimizes the communication for a given TSV count. For the objective $c_{util}$ (hop distance $\times$ bandwidth) the routing algorithm is used. A penalty is added for links with more load than bandwidth available ($c_{peak}$). Routers in adjacent layers with a distance smaller than the RD-length $R$ can be connected. The set of connections is chosen via a SA that randomly switches them.

\textbf{Step {\scriptsize\circled{5}}}: The solution is \textbf{legalized} to accommodate the area for added 3D routers and TSV arrays. The LP/SDP from the second step are used again.

The order of the five design steps is motivated as follows: For fast optimization, the step {\scriptsize\circled{1}} splits the design in layers before floor-planning each layer in the step  {\scriptsize\circled{2}}. Deciding the step {\scriptsize\circled{3}} as late as possible is beneficial for network design as then floor plans can be accounted for. This avoids overallocation of network resources. The steps {\scriptsize\circled{3}} and {\scriptsize\circled{4}} are separated to allow for comparison against standard approaches. Both the steps are interdependent and their separation is far from trivial; we opt for this to balance performance and accuracy. If desired, heuristic can be iterated for further improvement. 
\vspace{-2pt}
\section{Results}\label{sec:results}
\vspace{-2pt}

The heuristic is implemented in MATLAB. The LPs are solved with CPLEX 12.8.0 and the SDPs with Mosek 8.1. The source code is available at \url{https://github.com/jmjos/A-3D-NoC-DSE}. For the evaluation we use three case studies, based on a 3D Vision SoC (VSoC) \cite{Zarandy.2011}, a typical application for heterogeneity. The PPA for 45nm mixed-signal and the 28nm digital nodes are given in Tab.~\ref{tab:PPA}. ADCs must be implemented in a mixed-signal node. For the sake of generality, we only consider the relative differences of the power (measured in static and dynamic power) and the performance (measured in timing) generated from synthesis. The routing algorithm uses the shortest path. We assume face-to-back bonding in all cases.

\begin{table}
	\caption[]{Case Study PPA table.}\vspace{-6pt}
	\label{tab:PPA}
	{\scriptsize
		\begin{tabularx}{\linewidth}{X|r|r|r|r|r|r} 
			& \multicolumn{2}{r|}{Area [mm\textsuperscript{2}]} & \multicolumn{2}{r|}{Perf. [relative]} & \multicolumn{2}{r}{Power [relative]}\\
			Nodes: & \unit[28]{nm} & \unit[45]{nm} & \unit[28]{nm} & \unit[45]{nm}& \unit[28]{nm} & \unit[45]{nm}\\
			\midrule
			CPU (RISC-V) & 35.8 & 62.2 & 1 & 1.34 & 1 & 1.34 \\
			ADC \cite{Lyu.2014} & n.a. & 53 & n.a. & 1 & n.a. & 1\\
			SIMD (nu+) \cite{Flich.2018} & 71 & 125 & 1 & 1.34 & 1 & 1.34\\
			2D Router & 1.3 & 2.25 & 1 & 1.34 & 1 & 1.34 \\
			3D Router & 1.8 & 3.15 & 1 & 1.34 & 1 & 1.34 \\
			\bottomrule
			\multicolumn{7}{l}{KOZs: \unit[2]{mm\textsuperscript{2}}. Maximum length of RD: \unit[5]{mm}.}\\ 
	\end{tabularx}\vspace{-8pt}}
\end{table}

\textbf{Case study I:} The \textbf{tiny 3D SoC} has 5 CPUs and two \unit[28]{nm} digital layers. The application graph has bidirectional links with \unit[1]{Mb/s} bandwidth between subsequent nodes.  

\textbf{Case study II:} The \textbf{small 3D VSoC} has one \unit[28]{nm} digital layer one \unit[45]{nm}  mixed-signal layer. It implements 9 analog-digital converters (ADCs) and 9 CPUs.  It captures a 720p60 video stream, conducts AD-conversion and processes a convolution filter. In a conventional design, the ADCs are located in a $3\!\times\!3$ mesh NoC in the mixed-signal layer and the CPUs in a $3\!\times\!3$ mesh NoC in the digital layer. As RD is not used conventionally, the mesh sized in both layers are identical and routers are located at the same positions. 

\textbf{Case study III:} The \textbf{large 3D VSoC} implements 9 ADCs, 18 CPUs and 3 SIMD cores. The chip has one mixed-signal layer in \unit[45]{nm} node and two digital layers in \unit[28]{nm} node. The VSoC runs Viola-Jones, Shi and Tomasi and KLT algorithms for face recognition and tracking from a 720p60 video. In a conventional design, the ADCs are located in a $3\!\times\!3$ mesh NoC in the mixed-signal layer, the CPUs and SIMD cores in a $3\!\times\!3$ mesh NoC in the first digital layer and the remaining CPUs are in a $4\!\times\!3$ NoC below. 
%


\vspace{-3pt}
\subsection{Comparison to an optimal solution}\label{sec:discussion}
\vspace{-2pt}

\begin{table}
	\caption[]{Optimal MILP vs. Heuristic with LP and SDP.}\vspace{-6pt}
	\label{tab:optimal}
	{\scriptsize
		\begin{tabularx}{\linewidth}{X|r|r|r|r|r} 
			Method: &  \multicolumn{1}{c|}{Optimal} & \multicolumn{4}{c}{Heuristic} \\
			\textbf{tiny 3D SoC} & MILP & LP & $\Delta$& SDP & $\Delta$ \\
			\midrule
			Optimization Runtime [s] & 599 & 27.67& -95\% &152.4& -75\% \\
			Area [mm\textsuperscript{2}] -- upper layer & 158.0 & 158.0 & 0\% & 117.0 & -36\%\\ 
			Area [mm\textsuperscript{2}] -- lower layer & 77.4 & 77.4 & 0\% & 77.0 & -0.5\% \\
			Bandwidth$\times$Distance[mm\textsuperscript{2}Mb/s] & 18.51 & 31.08 & +41\%& 26.53& +15\% \\
			\bottomrule
			\multicolumn{6}{l}{Parameters: Step 2: initial temp. 20, 120 iters, 0.97 cooling; Step 4: initial temp. }\\
			\multicolumn{6}{l}{100, 50 iters, 0.97 cooling; Weights $\omega_i$ set to 1.}
	\end{tabularx}\vspace{-8pt}}
\end{table}

The \textbf{tiny SoC} provides a small example that can be solved optimally using an analytical mixed-integer linear model (MILP) to compare against the heuristic. 

We execute the heuristic using both the LP and the SDP to optimize area. The LP allows to compare against the MILP, while the SDP removes the linearization error. The results are shown in Tab.~\ref{tab:optimal}. Considering \emph{runtime}, the MILP is naturally very slow; it finds a solution within \unit[\textasciitilde9.5]{min}. The heuristic is much faster. Its efficient runtime with a linear model allows to use an SDP to remove the linearization error for area. The heuristic with SDP is still 75\% faster than the MILP. Considering \emph{area}, the MILP and heuristic with LP yield the same results as the LP used for area optimization in the SA in step {\scriptsize\circled{2}} uses a subset of the MILP's constraints. Thus, area is identical for the same floorplan, as given in this simple case. The SDP improves area by up to 36\%. Considering \emph{network performance}, the MILP has the best result as it does not separate interlayer and intralayer communication. The SDP outperforms the LP as a side-effect of better area. Regarding \emph{power and (component) performance}, the results of MILP and heuristic are equivalent, since both use the same LP.

\begin{table}
		\caption[Execution time of optimization]{Execution time of heuristic for random inputs.}\vspace{-6pt}
	\label{tab:executionTimeComplete}
	{\scriptsize
		\begin{tabularx}{\linewidth}{X|r|r|r|r} 
			\textsc{Heuristic's Part}&  \multicolumn{1}{c|}{{5 cmp.}} & \multicolumn{1}{c|}{{40 cmp.}} & \multicolumn{1}{c|}{{80 cmp.}}& \multicolumn{1}{c}{{1000 cmp.}}\\
			execution times &  \multicolumn{1}{c|}{{2 layers}} & \multicolumn{1}{c|}{{4 layers}} & \multicolumn{1}{c|}{{4 layers}}& \multicolumn{1}{c}{{3 layers}}\\
			\midrule
			{\tiny\circled{1}} {Comp.  to layer assignment} &0.4\:s &0.4\:s& 0.4\:s& 0.4\:s\\
			{\tiny\circled{2}} {Layer floor planning} &146\:s&375\:s &802\:s & 0.3\:h\\
			{\tiny\circled{3}} {Number of TSV arrays} &0.1\:s& 0.2\:s &0.2\:s & 5\:min\\
			{\tiny\circled{4}} {Placement of TSV arrays} &3.8\:s& 298\:s &1300\:s & 30\:h \\
			{\tiny\circled{5}} {Legalization} &0.2\:s& 0.5\:s  &0.5\:s&0.5\:s\\
			\midrule
			\textsc{complete} & 152\:s &675\:s&\cellcolor{tablegray}\textbf{2104\:s} & 31\:h\\
			\bottomrule
			\multicolumn{5}{l}{Parameters: Step 2: initial temp. 20, 120 iters, 0.97 cooling; Step 4: initial temp. }\\
			\multicolumn{5}{l}{100, 50 iters, 0.97 cooling; Weights $\omega_i$ set to 1. Intel i7-7740X, Ubuntu 16.04 LTS.}
	\end{tabularx}\vspace{-8pt}}
\end{table}

\vspace{-3pt}
\subsection{Heuristic run-time performance}
\vspace{-2pt}

Tab.~\ref{tab:executionTimeComplete} gives the exemplary runtime performance of the heuristic with SDP. A large realistic input set with 80 components and 4 layers finishes in 35\:min. The largest set suffers from the brute-force approach in step {\scriptsize\circled{3}}, but a method with higher performance can be found easily. The implementation of all steps is reasonably fast to optimize even very large inputs sets, so the runtime of the heuristic is adequate. 

\vspace{-3pt}
\subsection{Advantages of optimization with redistribution} 
\vspace{-2pt}

\begin{table}
		\caption[Comparison of RD-lengths]{Effect of RD on communication.}\vspace{-6pt}
	\label{tab:results:RDlength}
	
	\centering
	{\scriptsize
		\centering
		\begin{tabularx}{\linewidth}{Z||r|r|r} 
			\textbf{Small 3D VSoC}& \textsc{used RD}  & \multicolumn{2}{c}{\textsc{bandwidth$\times$distance}} \\
			\textsc{Maximum length of RD}& \textsc{length }[mm]  & \multicolumn{2}{c}{\textsc{in }[mm Mb/s]} \\
			\midrule
			0 mm (0\%) (\textsc{Baseline}) & 0& 47.78& \\
			90 mm (25\%) & 39.47&  44.57& -6.72\% \\
			180 mm (50\%) & 95.54 & 42.84& -10.34\% \\
			360 mm (100\%) &  354.36& 41.85&\cellcolor{tablegray} \textbf{-12.41\%} \\
			\midrule
			\textbf{Large 3D VSoC}& & \multicolumn{2}{c}{} \\
			\midrule
			0 mm (0\%) (\textsc{Baseline}) & 0& 46,49& \\
			90 mm (50\%) & 78.42 & 43.04& -7.42\% \\
			180 mm (100\%) & 131.39& 40.48& \cellcolor{tablegray}\textbf{-12.93\%} \\
			\bottomrule
			\multicolumn{4}{l}{Parameters: Step 2: initial temp. 30, 200 iters, 0.98 cooling; Step 4: initial temp. }\\
			\multicolumn{4}{l}{1000, 50 iters, 0.97 cooling; Weights $\omega_i$ set to 1, fixed number of TSVs.}
		\end{tabularx}\vspace{-8pt}
	}
\end{table}

RD allows connecting routers vertically that are not located exactly above each other. The enables more flexible vertical interconnections and thus increases the interconnection efficiency. For our experiments, the length of the RD is calculated from the vendor models and is reduced gradually to demonstrate the positive effect of RD on the application communication's hop distance. The RD decreases the hop distance by up to 12.93\%. For this improvement, the integrated approach is essential because both the floorplan and the horizontal NoC topology are required to find optimized vertical links for the application. Summarizing, RD improves the communication by \textasciitilde13\% over a baseline without RD.

\vspace{-3pt}
\subsection{Advantages of the integrated approach}
\vspace{-2pt}

\subsubsection{Whitespace reductions}

\begin{table}
		\caption[]{Conventional vs. optimized design.}\vspace{-6pt}
	\label{tab:results:heterogeneous}
	\centering
	{\scriptsize
		\centering
		\begin{tabularx}{\linewidth}{Z||r|r|r} 
			 Small 3D VSoC& \textsc{Baseline} & \textsc{Integrated} & \textsc{Difference}\\
			\midrule
			\textsc{Bandwidth$\times$distance} &\unit[2390]{mm\,Mb/s} &\unit[2830]{m\,Mb/s} & +18.41\%\\
			\textsc{Maximum link load} &\unit[120]{Mb/s}&\unit[190]{Mb/s}& +58.33\%\\
			\textsc{Whitespace} & \unit[30.48]{mm\textsuperscript{2}} & \unit[25.77]{mm\textsuperscript{2}} &\cellcolor{tablegray} \textbf{-15.44\%}\\
\midrule
			Large 3D VSoC & \textsc{Baseline} & \textsc{Integrated} & \textsc{Difference}\\
			\midrule
		    \textsc{Bandwidth$\times$distance} &\unit[2118]{mm\,Mb/s} &\unit[2599]{mm\,Mb/s} & +22.68\%\\
			\textsc{Maximum link load} &\unit[115.8]{Mb/s}&\unit[149.1]{Mb/s}& +28.77\%\\
			\textsc{Whitespace} & \unit[42.92]{mm\textsuperscript{2}} & \unit[34.86]{mm\textsuperscript{2}}  &\cellcolor{tablegray} \textbf{-18.79\%}\\
			\bottomrule
			\multicolumn{4}{l}{Parameters: Step 2 and 4: initial temp. 30, 500 iters, 0.97 cooling}\\
	\end{tabularx}\vspace{-8pt}
	}
\end{table}


The conventional design of the case studies is packaging-inefficient because of the size difference between ACDs, CPUs and SIMD cores. This yields whitespace. It is reduced by our integrated approach due to a higher degree of freedom in the placement of the components and the routers. The area and the network performance of conventional and optimized designs are given in Tab.~\ref{tab:results:heterogeneous}. 
We achieve 15.44\% and 18.79\% reductions in whitespace for the small and the large 3D VSoC, respectively. The algorithm does not consider interlayer communication and thus the communication is worse by 18-58\%. The results show the typical limitations of the conventional approach without RD and grid size variability. However, these features reduces the whitespace but, naturally, have a negative effect on the network performance. Whether this is an acceptable compromise depends on the design targets.

\subsubsection{Communication improvements}


\begin{table}
		\caption[NoC planning for heterogenenous 3D SoC]{Optimized network loads.}\vspace{-6pt}
	\label{tab:results:urand}
	\centering
	{\scriptsize
		\centering
		\begin{tabularx}{\linewidth}{Z||r|r|r} 
			{{Bandwidth$\times$distance}} [mm Mb/s] & \textsc{Baseline}& \textsc{Integrated} & $\Delta$\\
			\midrule
			\textsc{Small VSoC} & 10.46 & 10.46 & $\pm$0.00\% \\
			\textsc{Large VSoC} & 39.74 & 37.99 &\cellcolor{tablegray} \textbf{-4.40\%} \\
			
			\bottomrule
			\multicolumn{4}{l}{Parameters: Step 2: initial temp. 30, 200 iters, 0.97 cooling, step 4: initial temp.}\\
			\multicolumn{4}{l}{1000, 50 iters, 0.97 cooling; Weights $\omega_i$ set to 1, fixed number of TSVs.}\\
	\end{tabularx}\vspace{-8pt}
	}

\end{table}

To show the advantages of the integrated approach, we compare it against a solution without application information. Therefore, we use the cases' traffic vs.\ uniform random traffic, which does not provide application-specific traffic patterns. Tab.~\ref{tab:results:urand} gives the results. We do not see an improvement for the small VSoC, because the application's traffic is almost uniformly distributed. The communication is improved by 4.4\% for the large VSoC. 

\vspace{-3pt}
\subsection{Comparison against related work}
\vspace{-2pt}

\subsubsection{Component floor planning} We compare our floor plan against \cite{Srinivasan.2006}, as this work also accounts for different core sizes. We show different multimedia benchmarks in Tab.~\ref{tab:floorplanning}. For the first two benchmarks, we improve the area at costs of the communication. However, for the H263enc Mp3 dec benchmark we see a general improvement: We achieve up to 16.4\% better area and 2\% better communication using the integrated approach.

\begin{table}
	\caption[Execution time of optimization]{Comparison vs. Ref. \cite{Srinivasan.2006}.}\vspace{-6pt}
	\label{tab:floorplanning}
	{\scriptsize
		\begin{tabularx}{\linewidth}{X|r|r|r||r|r|r} 
			Benchmark &\multicolumn{3}{c||}{Area [mm\textsuperscript{2}]}&\multicolumn{3}{c}{BW$\times$Dist.[mm\textsuperscript{2}Mb/s]}\\
			& \cite{Srinivasan.2006} & SA & $\Delta$ & \cite{Srinivasan.2006}& SA & $\Delta$ \\
			\midrule
			H256decMp3dec &11301 &8244&-27.1\%&19858& 21280& +7.16\%\\
			mp3dencMp3dec &8568&8516&-0.61\%&17546&17572 & +.15\%\\
			H263encMp3dec &12535&10474&\cellcolor{lightgray}\textbf{-16.4}\%&255324 & 250187 & \cellcolor{lightgray}\textbf{-2\%}\\
			\bottomrule
			\multicolumn{7}{l}{Parameters: initial temp. 30, 15000 iters, 0.98 cooling; average over 30 reruns. }
	\end{tabularx}\vspace{-8pt}}
\end{table}

\subsubsection{Placement of 3D routers} 
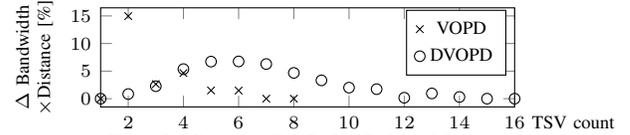
\begin{figure}
	\centering
	\begin{tikzpicture}
\begin{axis}[
height=2.9cm, width = .8\linewidth,
xlabel=TSV count,
ylabel={$\Delta$ Bandwidth\\$\times$Distance {[\%]}},
xmin = 1, xmax = 16 ,
xtick distance=2,
ylabel style={align=center, font=\scriptsize, yshift = -10},
xlabel style = {align=center, font=\scriptsize, at={(current axis.right of origin)}, ,anchor=west, xshift =3, yshift = 6},
tick label style={font=\scriptsize},
legend style = {font = \scriptsize},
legend = {} 
 ]
\addplot[only marks, mark=x] coordinates {
	(1,0)
	(2,14.97)
	(3,2.64)
	(4,4.64)
	(5,1.49)
	(6,1.47)
	(7,0)
	(8,0)
};

\addplot[only marks, mark=o] coordinates {
	(1,0)
	(2,.84)
	(3,2.26)
	(4,5.36)
	(5,6.72)
	(6,6.75)
	(7,6.28)
	(8,4.67)
	(9,3.32)
	(10,2.0)
	(11,1.73)
	(12,.17)
	(13,.95)
	(14,.31)
	(15,0)
	(16,0)
};
\legend{VOPD, DVOPD}
\end{axis}
\end{tikzpicture}
\vspace{-8pt}
\caption{SA vs. PSO \cite{Manna.2016} for third step.}
\label{fig:PSO}\vspace{-6pt}
\end{figure}

Using SA within our framework is justified by comparison against partial swarm optimization (PSO) \cite{Manna.2016} to determine the optimal number of TSV arrays. The difference in bandwidth$\times$distance is shown in Fig.~\ref{fig:PSO}. In average, SA is 3.125\% and 2.563\% better than PSO for Video Object Plane Detection (VOPD) and Double VOPD benchmarks, respectively. For 2 TSV arrays, SA is even 15\% better than PSO for the VOPD benchmark. These positive results make SA a reasonable method.

\vspace{-3pt}
\section{Summary}
\vspace{-1pt}
The system-level optimization improves the floor plan and the network topology of NoCs for heterogeneous 3D SoCs. Therefore, the proposed integrated approach considers properties of both the application and the technology. We propose a heuristic with five design steps for an efficient optimization that splits interlayer and intralayer communication. Whitespace is reduced by up to 18.8\%. Communication is better by up to 4.4\% through early consideration of traffic patterns.  
\vspace{-3pt}

\section*{Acknowledgments}
\vspace{-3pt}
\noindent This work is funded by DFG projects PI 447/8 and GA 763/7.
\vspace{-4pt}
\bibliographystyle{IEEEtran}
\bibliography{bibliographyShort}

\end{document}